\documentclass[11pt,a4paper]{article}
\usepackage{jheppub}
\usepackage{graphicx}
\usepackage{booktabs}
\usepackage{float}
\usepackage[caption = false]{subfig}
\usepackage{bm}

\newcommand{\be}{\begin{equation}}
\newcommand{\ee}{\end{equation}}
\newcommand{\bea}{\begin{eqnarray}}
\newcommand{\eea}{\end{eqnarray}}

\begin{document}

\title{Testing Pomeron flavour symmetry with diffractive W charge asymmetry}

\author[a]{A. Chuinard}
\author[b,c,d]{C. Royon}
\author[c]{R. Staszewski}

\affiliation[a]{McGill University, Montreal, Canada}
\affiliation[b]{Institute of Physics, Academy of Sciences of the Czech Republic, Prague, Czech Republic}
\affiliation[c]{Institute of Nuclear Physics, Polish Academy of Sciences, Krakow, Poland}
\affiliation[d]{The University of Kansas, Lawrence, USA}

\emailAdd{annabelle.chuinard@cern.ch}
\emailAdd{christophe.royon@cern.ch}
\emailAdd{rafal.staszewski@ifj.edu.pl}

\arxivnumber{1510.04218}

\date{\today}

\abstract{
This study focuses on hard diffractive events produced in proton-proton collision at LHC exhibiting one intact proton in the final state which can be tagged by forward detectors. We report prospective results on the W boson charge asymmetry measured for such events, which allow to constrain the quark diffractive density functions in the Pomeron.
}

\maketitle

\section{\label{sec:level1}Introduction}
As part of a theoretically challenging chapter of the Standard Model of particle physics, diffraction in quantum chromodynamics (QCD) is one of the thrilling topics in the physics programme at LHC. Right before the LHC era, the collaboration of high energy theorists and experimentalists from H1 and ZEUS at the Hoch Energie Ring Anlage (HERA), UA8 at CERN and D0 and CDF at the Tevatron contributed to major improvements of our knowledge of the hard diffractive part of the deep inelastic cross section~\cite{Boonekamp:2009yd}. This has been achieved by using perturbative QCD as a mean to resolve partonic sub-processes. Different approaches may be considered to model the physics at play in the different kinematic ranges.
However, experimental evidence from Fermilab's CDF proved that these approximations, although valid for Deep Inelastic Scattering (DIS), do not apply to diffractive interactions when it comes to hard hadron-hadron collisions~\cite{PRL2000}. A normalization factor (gap survival probability) must be introduced to compensate for the discrepancy between predicted and observed cross sections of diffractive processes. Measurements from ATLAS and CMS collaborations at LHC will allow to extract the value of the gap survival probability, often assumed to be constant, and to test the limits of the model~\cite{CERN-LHCC-2011-012}.

Of special interest is the diffractive gauge boson production for which the
CMS and CDF collaborations recently reported results based on rapidity gap
measurement~\cite{Chatrchyan:2011wb, Aaltonen:2010qe}.
The phenomenology of the single diffractive W production process has been
studied in the past, most often in the standard Ingelman-Schlein model of
diffraction~\cite{Ingelman:1984ns}. The focus was mainly on the the cross
section values and kinematic distributions~\cite{Ingelman:1984ns,Bruni:1993ym,Covolan:2002kh,GayDucati:2007irr,PhysRevD81}. In the present paper, we consider the charge asymmetry of single diffractive W's in the context of probing flavour structure of the Pomeron. We show
that this observable can be used to assess the mechanisms at play in hard
diffraction. At the LHC, the single-diffractive W production exhibits a
large enough cross-section to provide additional information on the
\mbox{"Pomeron-like"} picture considered for this study. The combined measurement
of the W and Z cross sections may also give insights on the Pomeron content,
but will not be explored in this paper.

\begin{figure}[htb!]
\centering
\includegraphics[width=5cm]{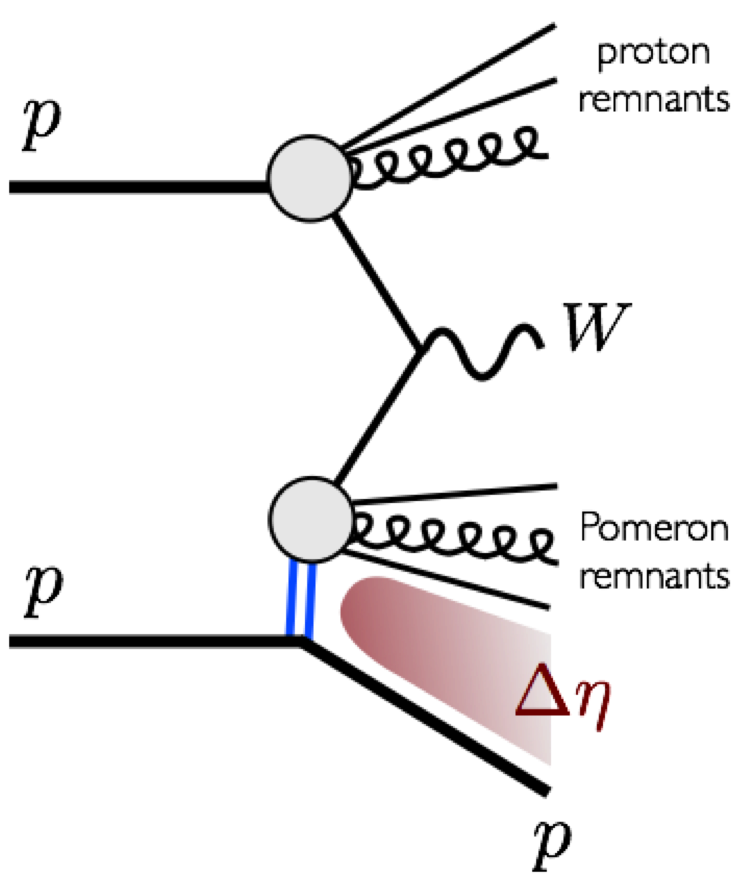}
\caption{Leading-order diagram for single-Pomeron exchange W boson production in proton-proton collisions. The double blue line corresponds to the colour singlet exchange. The  rapidity gap $\Delta \eta$ is showed as a faded pink area. The blobs denote the non-resolved structure of the colour singlet and that of the interacting proton.}
\label{fig:Fig1}
\end{figure}
In section \ref{sec:sectionII}, we present the salient features of the perturbative theory for hard diffraction in hadron-hadron collision, and their implementation into the Forward Physics Monte-Carlo generator (FPMC)~\cite{Boonekamp:2011ky} which is used for this analysis.
In section \ref{sec:sectionIII}, we discuss the possibility of using the charge asymmetry of  the W boson as a probe for the Pomeron structure. We focus on the quark content of the Pomeron and show which improvements can be obtained from such a measurement at LHC. Simulation results using FPMC are presented and discussed in section \ref{sec:sectionIV}. Section \ref{sec:sectionV} is dedicated to conclusion and outlook.

\section{\label{sec:sectionII} Hard single diffractive processes}

The study focuses on single diffractive W boson production in proton-proton collision at LHC i.e. $pp\to [W^{\pm} X]p$. The leading-order diagram for this process is pictured on Fig.\ref{fig:Fig1} and involves the exchange of a colour singlet object.

This process exhibits two characteristic features: a final state leading proton and a rapidity gap between this outgoing proton and the colour singlet object remnants where no particle is detected. Gap measurement and proton tagging are the two common techniques applied in collider experiments to select diffractive events. When possible, the results from both techniques are combined which allows to quantify the probability for the intact proton to dissociate into particles which cannot be tagged in the forward region.

In this study, we assume the intact proton to be measured in a forward detector. Compared to rapidity gap measurements, proton tagging is independent of the gap acceptance factors and of the proton dissociation effects. Thanks to a precise measurement of the proton and a good reconstruction of the primary vertex, this method can be used in an environment with a reasonably small level of pile-up, making it well suited for the second phase of the LHC "low mu" data taking. The recently approved ATLAS Forward Proton (AFP) and CMS-TOTEM Precision Proton Spectrometer (CT-PPS) foresee the installation of forward detectors using the LHC magnets as a spectrometer in order to tag and time very forward protons.

The AFP~\cite{Adamczyk:2017378} and CT-PPS~\cite{CERN-LHCC-2014-020, CERN-LHCC-2014-024, Albrow:1753795} projects comprise two detectors located at 203 m and 214 m either side of the interaction points (IP1 for ATLAS and IP5 for CMS). The ranges of energy and transverse momentum of the protons, often described by $\xi$ (the fraction of energy lost by the scattered hadron) and $t$ (the four momentum transfer squared), are particularly suitable for diffractive physics analysis~\cite{yellowreport} with AFP and CT-PPS. In addition, with a high $\beta^*$ optics, i.e. for a rather large beam spread at the interaction point, a good acceptance in $\xi$ (between 0 and 0.2) for moderate $t$ can be obtained with the vertical roman pot detectors (ALFA and CMS-TOTEM).

\subsection{Parton distribution functions}

The diffractive parton distributions are obtained from deep inelastic scattering (DIS) experiments using procedures similar to the ones used to derive proton PDFs. We shall consider however that, for the diffractive case, the system has additional degrees of freedom which must be integrated over in order to obtain the cross section. The diffractive component of the integrand is a function of the diffractive kinematics and may be expressed as a convolution of a Regge flux $\Phi_{\mathbb{P/R}}$ and a parton density function $f_{\mathbb{P/R}}$ describing the structure of the exchanged colour singlet object:
\be 
f(\xi,\beta, t, \mu^2) = \Phi_{\mathbb{P/R}} (\xi,t) \cdot f_{\mathbb{P/R}}(\beta,\mu^2)\label{eq:facto}
\ee
where indices $\mathbb{P}$ and $\mathbb{R}$ specify the type of trajectory considered, depending on the fraction of energy lost by the intact proton $\xi$ (sometimes referred as $x_\mathbb{P}$), as detailed in \mbox{section \ref{sec:xsec}}. In eq.\eqref{eq:facto}, $\beta$ denotes the fraction of the pomeron momentum carried by the interacting parton and $t$, the four-momentum squared transferred from the intact proton into the collision. The Reggeon flux is a function of the Regge trajectory and exhibits a dependence on $t$ as follows:
\be\Phi_{\mathbb{P/R}}=\frac{e^{B_{\mathbb{P/R}}t}}{\xi^{2\alpha_{\mathbb{P/R}}(t)-1}},
\label{eqflux}
\ee 

The diffractive slope $B_{\mathbb{P/R}}$ and the Regge trajectory $\alpha_{\mathbb{P/R}}(t)=\alpha_{\mathbb{P}}(0)+t\,\alpha ' _{\mathbb{P}}$ are calculated by fitting HERA data~\cite{Aktas:2006hy}. In the low $\xi$ regime, the Pomeron amplitude prevails over the Reggeon's (see \mbox{section \ref{sec:xsec}}). The parameters of the model are fitted, assuming a sub-leading exchange of Reggeons in addition to the Pomeron dominant contribution. The values obtained from HERA Fit B~\cite{Adloff:1997sc} and HERA FPS~\cite{Aktas:2006hx} are combined in Table \ref{tab:Tab2}. 
\begin{table}[H]
\caption{Diffractive structure function parameters of QCD used in FPMC. For the $\alpha_{\mathbb{P}}(0)$ term, the H1 DPDF Fit B from HERA~\cite{Adloff:1997sc} is used. $\alpha'_{\mathbb{P}}$ and $B_{\mathbb{P}}$ are obtained from H1 FPS data~\cite{Aktas:2006hx}.}
\centering
\begin{tabular}{cccc}
\toprule
Pomeron && Reggeon &\\
\midrule
$\alpha_{\mathbb{P}}(0)$&$1.111\pm{0.007}$&$\alpha_{\mathbb{R}}(0)$&$0.5\pm{0.10}$\\
$\alpha'_{\mathbb{P}}$&$0.06_{-0.06}^{+0.19}$ $\alpha'_{\mathbb{R}}$&$0.3_{-0.3}^{+0.6}$\\ 
$B_{\mathbb{P}}$&$5.5_{+0.7}^{-2.0}$&$B_{\mathbb{R}}$&$1.6_{+0.4}^{-1.6}$\\
\bottomrule
\end{tabular}
\label{tab:Tab2}
\end{table}

\subsection{Cross sections}\label{sec:xsec}

We derive the production cross section of W bosons at low $\xi$ from long-distance/short-distance factorization theorem adding the regularization term $S^2$ (gap survival probability) to account for the effects of the soft interactions that may smear out the rapidity gap and destroy the intact proton:
\begin{equation} 
\frac{d\sigma}{d\xi dt}= S^2 \sum_{i,j} \int dx_p \, d (\beta\xi) \, \mathcal{F}_{ij}(\xi, \beta, x_p, t, \mu^2) \,d\hat{\sigma}^{ij\to W} 
\end{equation}
with:
\[  \mathcal{F}_{ij}(\xi, \beta,x_p,t, \mu^2)=\underbrace{\Phi_{\mathbb{P}} (\xi,t)\cdot f_{\mathbb{P}}^i( \beta,\mu^2)}_\textrm{diffractive PDFs} \cdot \underbrace{ f_{p}^j(x_p,\mu^2)}_\textrm{ proton PDFs}\]
where $i$ and $j$ are indices running over the partons of the Pomeron and the hard proton respectively, and $d\hat{\sigma}^{ij \to W}$ is the differential partonic cross section for the hard production of the W boson.

\begin{figure}[htb!]
\centering
\includegraphics[width=0.65\linewidth]{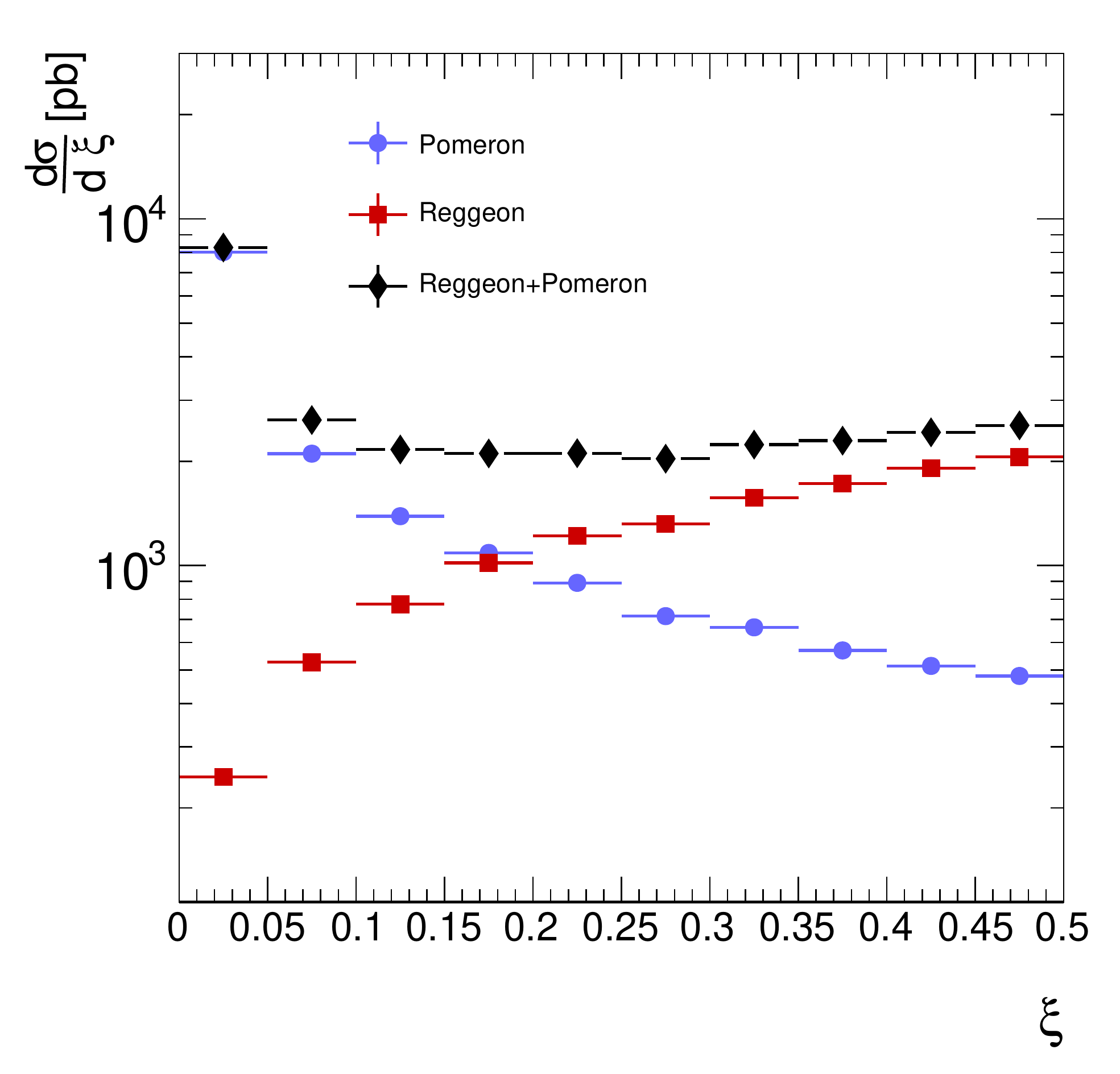}
\caption{Differential cross section as a function of $\xi$ for the production of single diffractive W bosons at $\sqrt{s}=14$ TeV given by the FPMC generator. Blue dots correspond to the Pomeron contribution, red squares the Reggeon contribution and black diamonds the sum of the two contributions.}
\label{fig:Fig6}
\end{figure}

Fig.\ref{fig:Fig6} shows the simulated inclusive cross section differential with respect to $\xi$ for W bosons produced via all the partonic and leptonic channels at a centre-of-mass energy of 14 TeV. $S^2$ is chosen to be constant and equal to 0.03 although its value may vary as we increase the centre-of-mass energy or change the process considered~\cite{Khoze:2000wk}. 

In the high $\xi$ regime, $\xi >0.2$, the Reggeon contribution dominates and adds up to the Pomeron contribution as follows:
\be
 \Phi_{\mathbb{P}} (\xi,t) \cdot f_{\mathbb{P}}^i(\beta ,\mu^2)+\Phi_{\mathbb{R}} (\xi,t) \cdot f_{\mathbb{R}}^i(\beta,\mu^2)  \label{eq:regpom}
\ee

Since the Pomeron contribution is high for large values of $\xi$ (see Fig.\ref{fig:Fig6}), we may consider adding a coupling term to eq.(\ref{eq:regpom}) to account for the possible interference between the Pomeron and Reggeon amplitudes. However, there is no evidence from the previous HERA investigations that interference plays a role in processes involving a colour singlet exchange. The fits assuming either minimal or maximal interference are both in good agreement with the data~\cite{Adloff:1997sc}. As a matter of simplicity, this study considers the interference term to be 0.

The results presented at \mbox{section \ref{sec:sectionIV}} only consider the regime at low $\xi$ where the Pomeron contribution is dominant compared to the Reggeon i.e. $\xi <0.12$. Other approaches to hard diffraction are also possible. Most interestingly, theories based on soft exchanges involving a change in the colour flow \cite{Edin:1995gi,Edin:1996mw,Rathsman:1998tp} can exhibit diffractive signatures, including the diffractive production of electroweak boson, without drawing on the concept of Pomeron \cite{Ingelman:2012wj}.
\begin{table}[H]
\centering
\caption{Cross sections for production of single-diffractive W bosons through the muonic channel $ W \to \mu \nu_\mu$ given by the FPMC generator. Both Pomeron and Reggeon pictures are presented with the centre-of-mass energy of the collision ranging from 2 to 14 TeV with full or limited detector acceptance, with no gap survival correction ($S^2=1$).}
\label{tab:table1}
\begin{tabular}{ccccc}
\toprule
Acceptance  	&$\sqrt{s}$	& $\sigma_\mathbb{P}$ [pb]&  	 $\sigma_\mathbb{R}$ [pb]&  $\sigma_\mathbb{P}/\sigma_\mathbb{R}$\\
 \midrule
%

$0<\xi<1$ 	& 2 TeV		    &	251.8		&	1089	& 0.231	\\
            &7 TeV	        &	1492		&	4021	& 0.371	\\
            &13 TeV		    &	3387		&	7200    & 0.470	\\
            &14 TeV	        & 	3698		&	7838    & 0.472
\\
            \midrule
$0.02<\xi<0.12$ & 2 TeV		&	68.10		&		19.69& 3.46	\\
				& 7 TeV		&	493.1		&		91.35& 5.40	\\       
         		&13 TeV		&	1213		&		177.8&6.82	\\
            	&14 TeV	    & 	1350		&		188.3&	7.17\\ 
\bottomrule
\end{tabular}
\label{xsectab}
\end{table}
\section{\label{sec:sectionIII} Charge asymmetry as a probe for Pomeron partonic structure}
\subsection{Implementation of the parton structure}
Both Pomeron and Reggeon pictures are implemented in FPMC. This is done essentially by substituting the proton density function by its diffractive counterpart. The HERWIG matrix elements of inelastic production are used to calculate the partonic cross sections at leading order.  
Table \ref{tab:table1} summarizes the cross sections predicted by FPMC for W bosons produced via a Pomeron-induced ($\sigma_\mathbb{P}$) or Reggeon-induced ($\sigma_\mathbb{R}$) single diffractive process and decaying through the  channel $ W \to \mu \nu_\mu$ for various values of the centre-of-mass energy. We present the results for a value of the gap survival probability which corresponds to no correction ($S^2=1.00$). Since the gap survival probability is changing with the energy scale, the displayed cross sections should be rescaled to correspond to the physics expectations at each energy level. For higher energies, the contribution of the Pomeron increases with $\sqrt{s}$ as reflected by the increase of the gap-survival-independent ratio of the  cross sections $\sigma_\mathbb{P}/\sigma_\mathbb{R}$.\\
This study fits in line with the HERA DIS measurement of the DPDFs considering the Pomeron to be composed of quarks and gluons. In such a picture, the charge neutrality of the Pomeron is accounted for by imposing symmetry conditions on the DPDFs introduced in \mbox{section \ref{sec:xsec} : $u_\mathbb{P}=\bar{u}_\mathbb{P}$, $d_\mathbb{P}=\bar{d}_\mathbb{P}$ and $s_\mathbb{P}=\bar{s}_\mathbb{P}$}. The differential cross sections are expected to be sensitive to the diffractive structure function $F_2^{D(4)}$, defined analogously to the inclusive proton structure function $F_2$. Since HERA inclusive measurements on DIS cross sections were non flavour-sensitive, the DPDFs were all assumed to be the same: $u_\mathbb{P}=d_\mathbb{P}=s_\mathbb{P}\equiv q$~\cite{GolecBiernat:2011dz}. Releasing this constraint introduces other sets of DPDFs that still have to satisfy this sensitivity requirement to the first order~\cite{Breitweg:1997aa} :
\be F_2^{D(4)} \propto \left( \frac{2}{3}\right)^2u_\mathbb{P} + \left( \frac{1}{3}\right)^2 d_\mathbb{P}+ \left(- \frac{1}{3}\right)^2s_\mathbb{P} \label{eq:F2D}\ee

Let $q$ be the HERA-derived quark distribution, then the equation to be fulfilled in order to satisfy \eqref{eq:F2D} is:
\[ 4u_\mathbb{P}+d_\mathbb{P}+s_\mathbb{P} = 6q
\] 
The solution to this equation is completely defined by two parameters. One choice of parameters is: 
\[ R_{ud}= \frac{u_\mathbb{P}}{d_\mathbb{P}}, \, \, \, R_{sd}=\frac{s_\mathbb{P}}{d_\mathbb{P}}
\]
The DPDFs can then be expressed as follows:
\begin{align}
u_\mathbb{P}(\beta, \mu^2) &= \frac{6R_{ud}}{1+R_{sd}+4R_{ud}}\cdot q(\beta, \mu^2) \label{eq:u}\\
d_\mathbb{P}(\beta, \mu^2) &= \frac{6}{1+R_{sd}+4R_{ud}}\cdot q(\beta, \mu^2)\label{eq:d}\\
s_\mathbb{P}(\beta, \mu^2) &= \frac{6R_{sd}}{1+R_{sd}+4R_{ud}}\cdot q(\beta, \mu^2)\label{eq:s}
\end{align}\\

By setting $R_{ud}=R_{sd}=1$, the default HERA distribution is obtained.
The above scaling factors are implemented into FPMC. Samples with values of $R_{ud}$ and $R_{sd}$  equal to 0.5, 1.0 and 2.0 were produced. One should keep in mind that, although not explicitly stated here, the DPDFs ratios may depend on the value of $\beta$ and $\mu^2$. For the purpose of this study, $R_{ud}$ and $R_{sd}$ are assumed to be constant.

\subsection{HERA constraints on the PDF ratios}\label{sec:SectionC}

We extract constraints on the Pomeron PDFs from previous HERA DIS diffractive charged current measurement in $ep$ collisions~\cite{Aktas:2006hy}: $e^- +q \to \nu +q'$.  In this case, a $W$ boson is exchanged in the $t$-channel that couples to the quarks of the Pomeron. The cross section for such a process is therefore sensitive to the quark content of the Pomeron and can be expressed as:
\begin{align}
\sigma \sim \int \text{d}x \, \text{d}Q^2 
&\Big( 
 u_{\mathbb{P}} \vert V_{ud} \vert^2 
+u_{\mathbb{P}} \vert V_{us} \vert^2 \nonumber \\
&+d_{\mathbb{P}} \vert V_{ud} \vert ^2
+s_{\mathbb{P}} \vert V_{us} \vert ^2 \nonumber \\
&+\rho(Q^2) s_{\mathbb{P}} \vert V_{sc} \vert^2
+\rho(Q^2) d_{\mathbb{P}}\vert V_{dc} \vert^2
\Big)\label{eq:sigma_H1}
\end{align}
\noindent where $ u_{\mathbb{P}}= \bar{u}_{\mathbb{P}}$, $ d_{\mathbb{P}}=\bar{d}_{\mathbb{P}}$ and  $s_{\mathbb{P}}=\bar{s}_{\mathbb{P}}$ are the DPDFs of the Pomeron which depend on the energy scale $Q^2$ and the Bj\"{o}rken scaling variable $x$, $V_{qq'}$ corresponds to the CKM matrix element coupling quark $q$ to quark $q'$. The suppression of the contributions involving heavy quarks is accounted for by adding an energy dependent correction factor $\rho(Q^2)$ ranging from 0 (full suppression) to 1 (no suppression) to the terms that involve $c$ quark interaction. Using eq.(\ref{eq:u}), eq.(\ref{eq:d}) and eq.(\ref{eq:s}) for the DPDFs, we explicit eq.(\ref{eq:sigma_H1}) as follows:
\begin{align*}
\sigma = A \cdot
\frac{1}{1+R_{sd}+4R_{ud}}&\Big[
R_{ud} \big(\vert V_{ud}\vert ^2 + \vert V_{us}\vert ^2 \big)+ \vert V_{ud} \vert^2 + \bar\rho \vert V_{dc} \vert^2 + R_{sd} \big(\vert V_{us} \vert^2 + \bar\rho \vert V_{sc} \vert ^2 \big)\Big]
\end{align*}
where $\bar{\rho}$ is the average $\rho$ over $Q^2$, and 
\[
A\sim \int q(x, Q^2)\, \text{d}x \, \text{d}Q^2 
\]

\begin{figure}[H]
\centering
\includegraphics[width=0.55\linewidth]{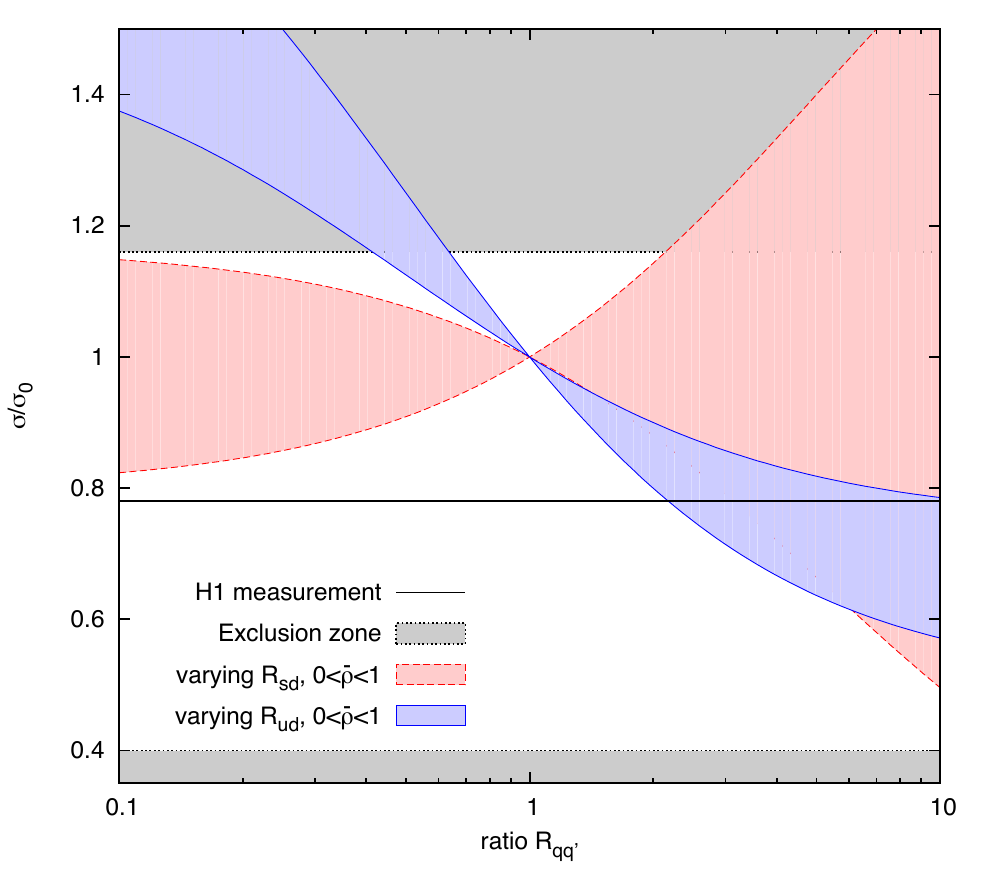}
\caption{Constraints on the quark content of the Pomeron. $\sigma_0$ (resp. $\sigma$) corresponds to the cross section with (resp. without) the $R_{ud}=R_{sd}=1$ hypothesis. The black solid line is the value measured by H1. The excluded region defined in eq.\ref{eq:eqcr} appears as a black dotted lines. The filled area delimited by blue plain (resp. red dashed) lines corresponds to the case were $R_{ud}$ (resp. $R_{sd}$) is varied for $\bar\rho$ ranging from 0 to 1.}
\label{fig:Fig8}
\end{figure}

The cross section for this process when assuming the equality of the DPDFs is predicted to be \mbox{$\sigma_0 = \sigma \rvert_{R_{ud}=R_{sd}=1}= 500\ \text{fb}$}, while the actual H1 measurement~\cite{Aktas:2006hy} gave: $\sigma = 390 \pm 120 \, \text{(stat.)} \pm 70 \,\text{(syst.) fb}$. These values can be used to constrain the ratio of the cross sections as: \begin{equation}0.4<\sigma/\sigma_0<1.16\label{eq:eqcr}\end{equation}
where\begin{align*}
\sigma/\sigma_0 =&  \frac{1}{2\vert V_{ud}\vert ^2 + 2\vert V_{us} \vert^2+ \bar\rho \vert V_{dc} \vert^2+ \bar\rho \vert V_{sc}\vert^2}\cdot \frac{6}{1+R_{sd}+4R_{ud}}\\
\cdot &\bigg[ R_{ud} \Big(\vert V_{ud} \vert^2+ \vert V_{us}\vert^2\Big) 
+ \vert V_{us}\vert^2 + \bar\rho \vert V_{sc} \vert^2 + R_{sd} \Big(\vert V_{ud}\vert^2 + \bar\rho \vert V_{dc}\vert^2\Big)
\bigg].
\end{align*}
Fig.\ref{fig:Fig8} shows the limits the H1 measurement sets on the flavour ratios. We deduce that $R_{ud}$ must satisfy $0.4<R_{ud}$, and that there is no constraint on $R_{sd}$.  For the purpose of this study, we focus on 6 cases in agreement with Fig.\ref{fig:Fig8}: $R_{ud}=0.5,\,1.0,\,2.0$ and $R_{sd}=0.5, \, 1.0,\,2.0$.
\begin{figure}[H]
\centering
\includegraphics[width=0.48\linewidth]{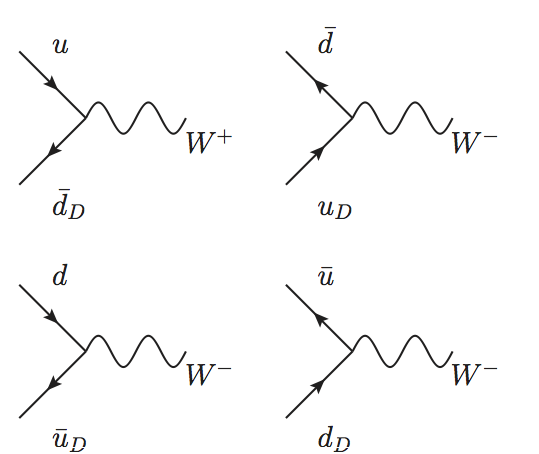}
\caption{Leading-order diagrams for W boson production. The CKM suppressed $[us]$ vertices are not shown here. For $pp$ collisions, the sea quarks contributions, as depicted on the right of the figure, are suppressed.}
\label{fig:Fig2}
\end{figure}

\subsection{Charge asymmetry}
Provided a large enough transverse momentum of the W boson, the hard cross section $d \hat{\sigma}^{ij\to W}$ can be approximated at first order by:
\be
d\sigma^{ij\to W}=\frac{2\pi G_F}{3\sqrt{2}}M_W^2 \vert V_{ij}\vert^2 \delta(\hat{s}-M_W^2) \label{eq:hardxs}
\ee
where $M_W$ is the mass of the W boson, $G_F$ is the Fermi constant and $V_{ij}$ is the CKM matrix element associated to quarks $i$ and $j$. The $ud$ vertices are depicted on Fig.\ref{fig:Fig2}. Assuming that we are in a $\xi$ regime that allows the Reggeon vertices to be neglected ($\xi<0.12$, see \mbox{section \ref{sec:xsec}}), we derive the following relation for the differential cross section as a function of the rapidity $y$ of the $W$ boson:

\be
\frac{d\sigma}{dy \,d\xi}= S^2 G_0 \sum_{ij} \vert V_{ij} \vert^2 \, \Phi_\mathbb{P}(\xi,t)\, f_\mathbb{P}^i(\beta, M_W^2) \, f_p^j(x_p,M_W^2)  \label{eq: diffxseccharge}
\ee
where $G_0=\frac{2\pi G_F M_W^2}{3\sqrt{2}s}$.\\

In terms of the two charges we get:
\begin{align*}
\frac{d\sigma_{W^+}}{dy\, d\xi}= S^2 \, G_0 \, \Phi_\mathbb{P} \Big( &\vert V_{ud} \vert^2 \left[u_\mathbb{P} \cdot \bar{d}_p+ \bar{d}_\mathbb{P}\cdot u_p\right]\\\nonumber +&\vert V_{us}\vert^2 \left[\bar{s}_\mathbb{P}\cdot u_p+u_\mathbb{P}\cdot \bar{s}_p\right] \Big)\\
\frac{d\sigma_{W^-}}{dy \, d\xi}= S^2 \, G_0 \, \Phi_\mathbb{P} \Big(& \vert V_{ud} \vert^2 \left[\bar{u}_\mathbb{P}\cdot d_p + d_\mathbb{P} \bar{u}_p\right] \\+&\vert V_{us}\vert^2 \left[ s_\mathbb{P}\cdot \bar{u}_p +\bar{u}_\mathbb{P}\cdot s_p \right] \Big) 
\end{align*}
where the $\beta, \xi$ and $t$ dependences of the Pomeron flux and PDFs are not displayed here for simplicity.\\

The charge asymmetry $\mathcal{A} $ of the W bosons produced through a single diffractive mechanism is defined as the normalized difference between the cross sections of production of $W^+$ and $W^-$. It is measured in terms of the average number of positive and negative Ws, $N^+$ and $N^-$, for a given integrated luminosity:
\be 
\mathcal{A}= \frac{d\sigma_{W^+}  -d\sigma_{W^-} }{d\sigma_{W^+} +d\sigma_{W^-}} \sim \frac{N^+-N^-}{N^++N^-}  \label{eq:asym}
\ee
$\mathcal{A}$ is a particularly good observable since it is not sensitive to the gap survival probability which cancels out when dividing numerator and denominator. Another advantage is that the systematic contributions to the uncertainty that are common to $W^+$ and $W^-$ measurements will cancel out when taking the ratio, which allows for more precision compared with a study relying  solely on the cross sections. Combining \eqref{eq:u}, \eqref{eq:d},  \eqref{eq:s}, \eqref{eq: diffxseccharge} and \eqref{eq:asym}, we have:
\begin{equation} 
\frac{1-\mathcal{A}}{1+\mathcal{A}}=\frac{\vert V_{ud} \vert^2 \left[ R_{ud}\cdot d_p + \bar{u}_p \right] + \vert V_{us} \vert^2 R_{sd} \left[ \bar{u}_p+s_p\right]}{\vert V_{ud} \vert^2 \left[ R_{ud}\cdot \bar{d}_p + u_p \right] +  \vert V_{us} \vert^2  R_{sd} \left[ u_p+ \bar{s}_p\right]}
\label{minuseq}
\end{equation}
displaying an explicit dependence on the $R_{ud}$ and $R_{sd}$ Pomeron flavour ratios. Relation \ref{minuseq} can also be written in terms of of the valence quark $q^v$ and see quark $q^s$ distributions:
\begin{equation*} 
\frac{\vert V_{ud} \vert^2 \left[ R_{ud}\cdot (d_p^v+d_p^s) + \bar{u}_p^s \right] 
+ \vert V_{us} \vert^2 R_{sd} \left[ \bar{u}_p^s+s_p^v+s_p^s\right]}
{\vert V_{ud} \vert^2 \left[ R_{ud}\cdot \bar{d}_p^s + u_p^v +u_p^s \right]
 + \vert V_{us} \vert^2  R_{sd} \left[ u_p^v+u_p^s+ \bar{s}_p^s\right]}
\end{equation*}
When $x_p>\xi$ (large rapidities), the sea quarks contribution can be neglected and we obtain the following equation implying the proton valence quarks:
\begin{equation*} 
\frac{1-\mathcal{A}}{1+\mathcal{A}}\rightarrow \frac{\vert V_{ud} \vert^2  R_{ud}\cdot d_p^v 
+ \vert V_{us} \vert^2 R_{sd} \cdot s_p^v}
{\vert V_{ud} \vert^2  R_{ud}\cdot u_p^v
 + \vert V_{us} \vert^2  R_{sd} \cdot u_p^v}
\end{equation*}

\section{Prospective results using FPMC}\label{sec:sectionIV}

\begin{figure}[H]
\centering\includegraphics[width=0.65\linewidth]{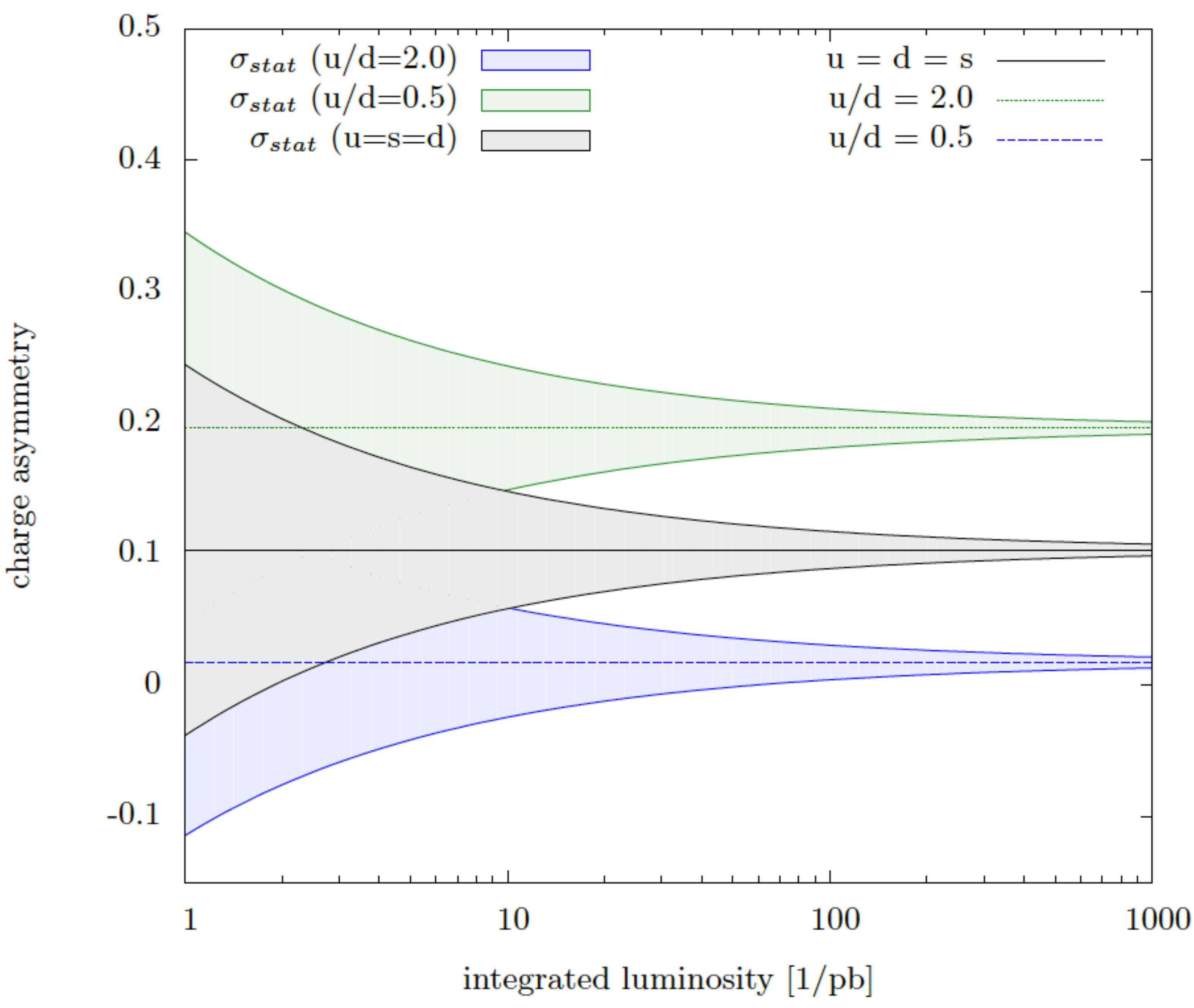}
\caption{Statistical uncertainty for various integrated luminosity for each $R_{ud}$ ratio for samples with $\xi<0.12$.}
\label{fig:Fig7}
\end{figure}

\begin{figure}[H]
\centering
\captionsetup[subfloat]{farskip=2pt,captionskip=1pt}
\subfloat[Intact proton momentum loss fraction]{\centering \includegraphics[width=.32\linewidth]{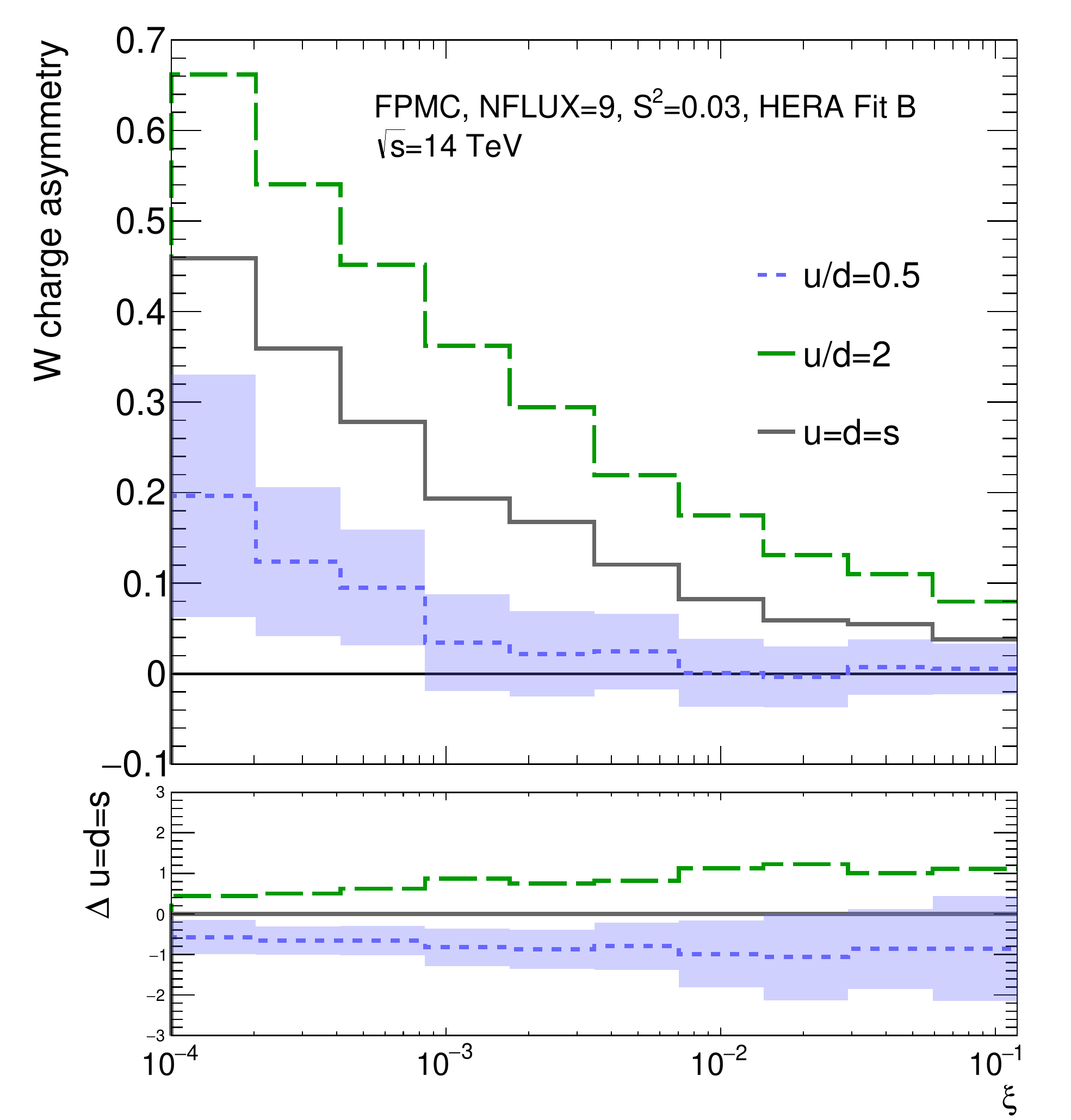}\label{fig:Fig3a}}
\hfill
\subfloat[W rapidity]{\centering \includegraphics[width=.32\linewidth]{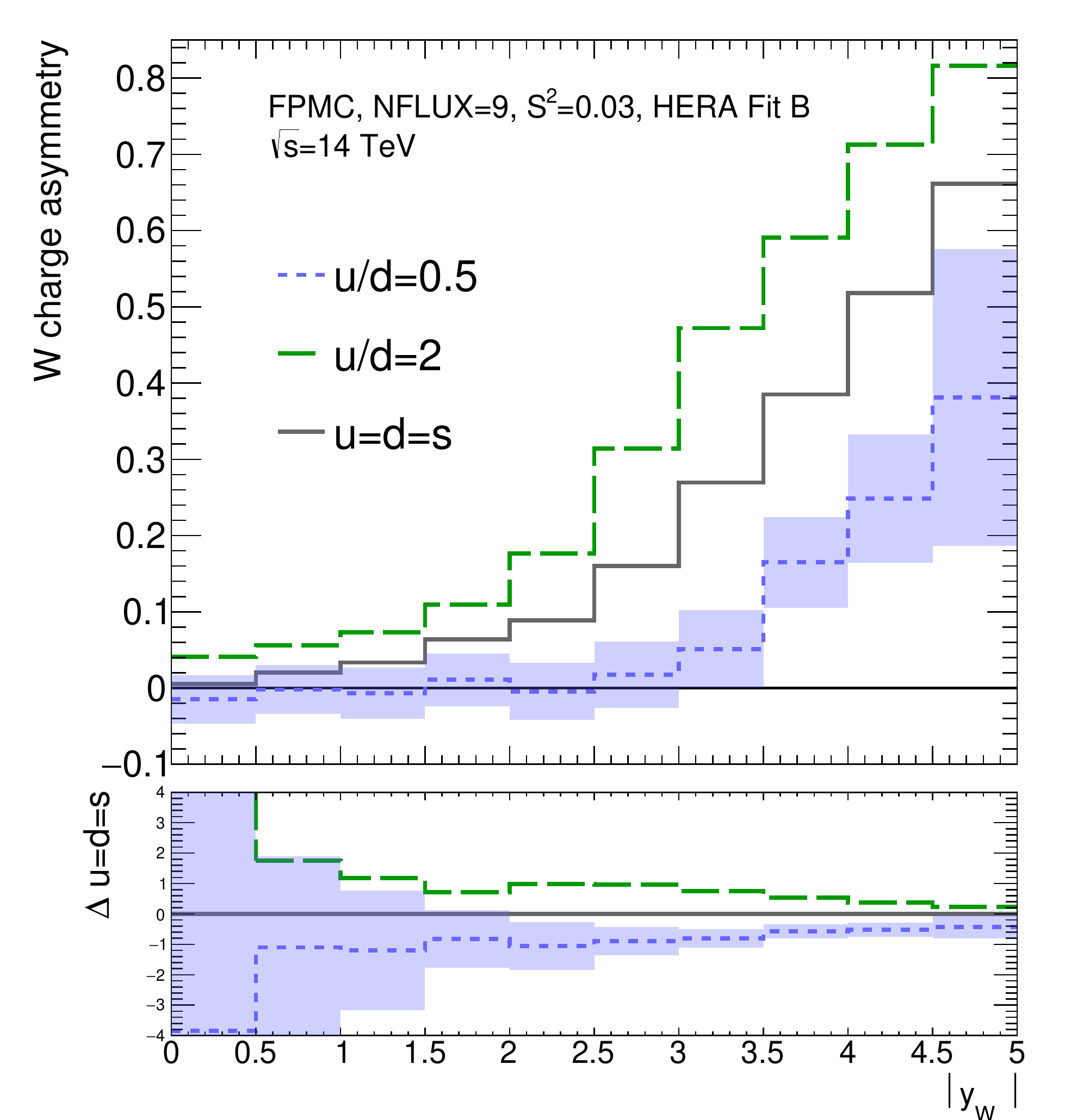}\label{fig:Fig3b}}
\hfill
\subfloat[Interacting fraction of Pomeron momentum]{\centering \includegraphics[width=.32\linewidth]{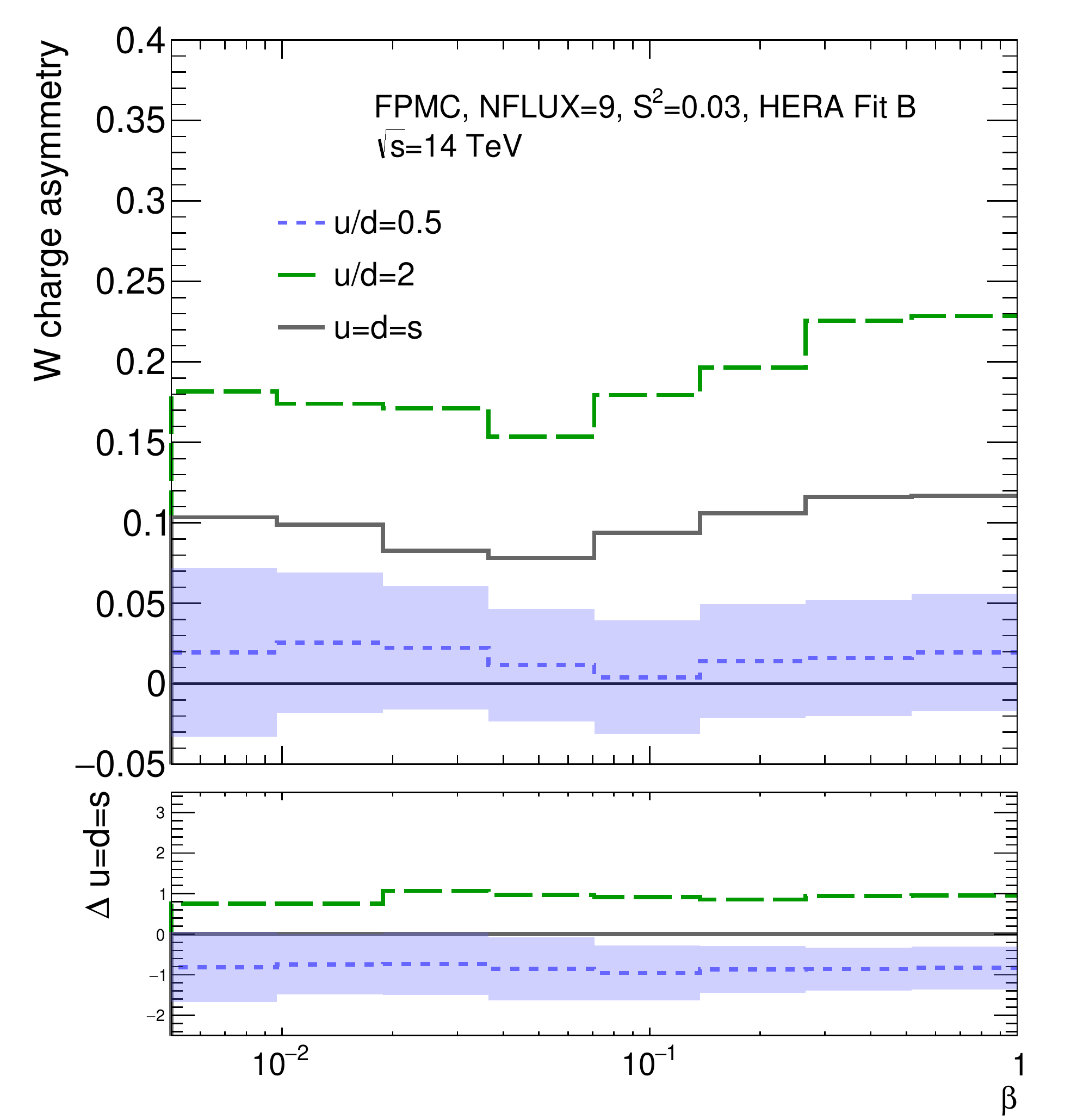}\label{fig:Fig3c}}\\
\caption{W charge asymmetry and its relative deviation from the case of equality of the DPDFs (black line) as a function of $\xi$(5a), $y_W$(5b) and $\beta$(5c) for various values of $R_{ud}$ and $\xi<0.12$. The statistical uncertainty associated to a 100 pb$^{-1}$ integrated luminosity is drawn in blue for the $R_{ud}=0.5$ hypothesis.}
\label{fig:Fig3}
\end{figure}

The following results are obtained by simulation of $pp\to [W^{\pm}X]p$ at the LHC energy $\sqrt{s}=14$ TeV with FPMC using the theoretical framework described in \mbox{sections \ref{sec:sectionII} and \ref{sec:sectionIII}}, assuming a Pomeron-like object and fitting the PDFs with HERA Fit B. As mentioned in \mbox{section \ref{sec:sectionII}}, the intact protons are required to have $\xi<0.12$ such that the Reggeon contribution can be neglected as well as the Pomeron-Reggeon interference. \\
The statistical uncertainty calculated at various luminosity levels for each hypothesis on the $R_{ud}$ ratio is presented on Fig.~\ref{fig:Fig7}. A minimum of 10 pb$^{-1}$ is required in order to be able to distinguish between the different hypothesis when considering the integrated asymmetry. We see that we can constrain the quark distributions in the Pomeron and study if they are equal. In Fig.\ref{fig:Fig3}, we present the dependence of the W charge asymmetry over a range of kinematic variables: $\xi$ (\ref{fig:Fig3a}), $y_W$(\ref{fig:Fig3b}) and $\beta$ (\ref{fig:Fig3c}) for FPMC samples generated with different DPDFs ratio.
The exchange of a Pomeron-like object is assumed. The $R_{sd}=0.5$ and $R_{sd}=2.0$ cases are not displayed because the distributions were found to be compatible with the $u_{\mathbb{P}}=s_{\mathbb{P}}=d_{\mathbb{P}}$ hypothesis.
Statistical uncertainties are shown for an expected integrated luminosity of 100 $\textrm{pb}^{-1}$ for $R_{ud}=0.5$ only. The size of the error bars is very similar for the two other cases. 
The three bottom plots show the relative deviation $\Delta_{u=d=s}$ from the asymmetry imposed by the HERA hypothesis $\mathcal{A}_H$:
\begin{equation*} 
\Delta_{u=d=s}= \frac{\mathcal{A}-\mathcal{A}_H}{\mathcal{A}_H}
\end{equation*}

On Fig.\ref{fig:Fig3b} we show the spectrum for absolute values of the rapidity:
\be \vert y_W \vert = \left \vert \frac{1}{2}\ln\left( \frac{E_W+p_{z,W}}{E_W-p_{z,W}} \right) \right \vert
\label{eq:rapidity}
\ee

\begin{figure}[h]
\centering
\includegraphics[width=0.42\linewidth]{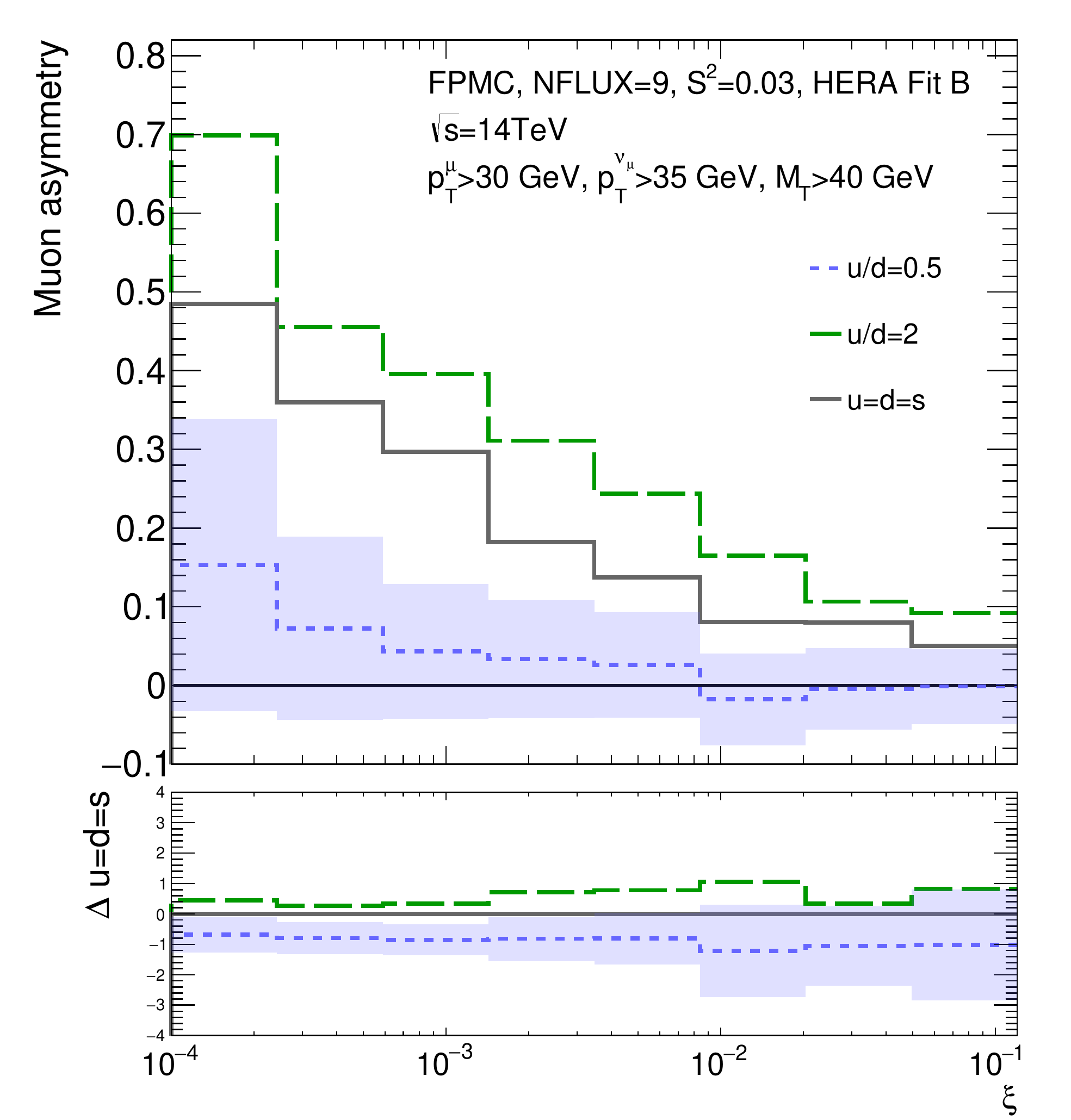}
\caption{Charge asymmetry of single-diffractive muons produced through the channel $W\to \mu \nu_\mu$ as a function of $\xi$ for various values of $R_{ud}$. The black line denotes the HERA hypothesis of equality of the DPDFs. The error bars represents statistical uncertainty for $R_{ud}=0.5$ scaled to the expected cross section at 14 TeV with $L=300\,\textrm{pb}^{-1}$. The error bars for the other hypothesis show little difference and are not represented. The bottom plots show the relativistic deviation from the $u_{\mathbb{P}}=s_{\mathbb{P}}=s_{\mathbb{P}}$ hypothesis.}
\label{fig:Fig4}
\end{figure}

We observe no sensitivity to the $R_{sd}$ ratio. However we see discrepancies between the $R_{ud}=0.5$ (blue dotted line) and $R_{ud}=2$ (green dashed line) cases.

In Fig.\ref{fig:Fig3c}, we present the asymmetry spectrum as a function of the fraction of the Pomeron momentum carried by the interacting parton $\beta$. The values of $\beta$ are extracted from the Monte-Carlo observables by means of formula \eqref{eq:rapidity} and using $M_W= \sqrt{\xi \beta x_p s}$ for the mass of the W boson:
\be
\beta= \frac{1}{\xi}\frac{M_W}{\sqrt{s}} e^{-y_W} 
\ee 
A similar trend is observed on Fig.\ref{fig:Fig4} which shows the charge asymmetry for the muons from the $W\to \mu \nu_\mu$ decay. The muon samples were generated with FPMC recording the decay products of the W. The signal is obtained from the detection of the muon and the measurement of the neutrino via the missing transverse energy of the system.
Most of the QCD background may be removed by applying a set of standard selection criteria for leptonic decays listed above. The selection presented here is specific to ATLAS but similar cuts may be used to select the single diffractive W bosons in the CMS data:
\begin{itemize}
\item Transverse momentum of the muon: $p_T^\mu >30$ GeV,
\item Missing transverse energy: $p_T^{\nu_\mu} >35$ GeV,
\item Transverse mass of the W boson:\[M_T= \sqrt{(E_{T,\mu}+E_{T,\nu_\mu})^2-(\vec{p}_{T,\mu}+\vec{p}_{T,\nu_\mu})^2}>40\textrm{ GeV}\] where $E_{T,\mu}=\sqrt{m_\mu^2+\vec{p_{T,\mu}}^2}$ and $E_{T,\nu_{\mu}}=p_{T,\nu_\mu}$ are the transverse energies of the muons and the neutrino and $\vec{p}_{T,\mu}$ and $\vec{p}_{T,\nu_\mu}$ their transverse momenta.
\end{itemize}

We consider only statistical uncertainties at $\sqrt{s}=14$ TeV for an integrated luminosity of 300 $\textrm{pb}^{-1}$. Only the $R_{ud}=0.5$ error bars are displayed to ease the reading. The amplitudes for the other cases are very similar. As in Fig.\ref{fig:Fig3a}, sensible deviation is observed with respect to the $R_{ud}=R_{sd}=1$ scenario.

\begin{figure}[H]
\centering
\captionsetup[subfloat]{farskip=2pt,captionskip=1pt}
\subfloat[Muon $p_T$ cuts]{\centering \includegraphics[width=.32\linewidth]{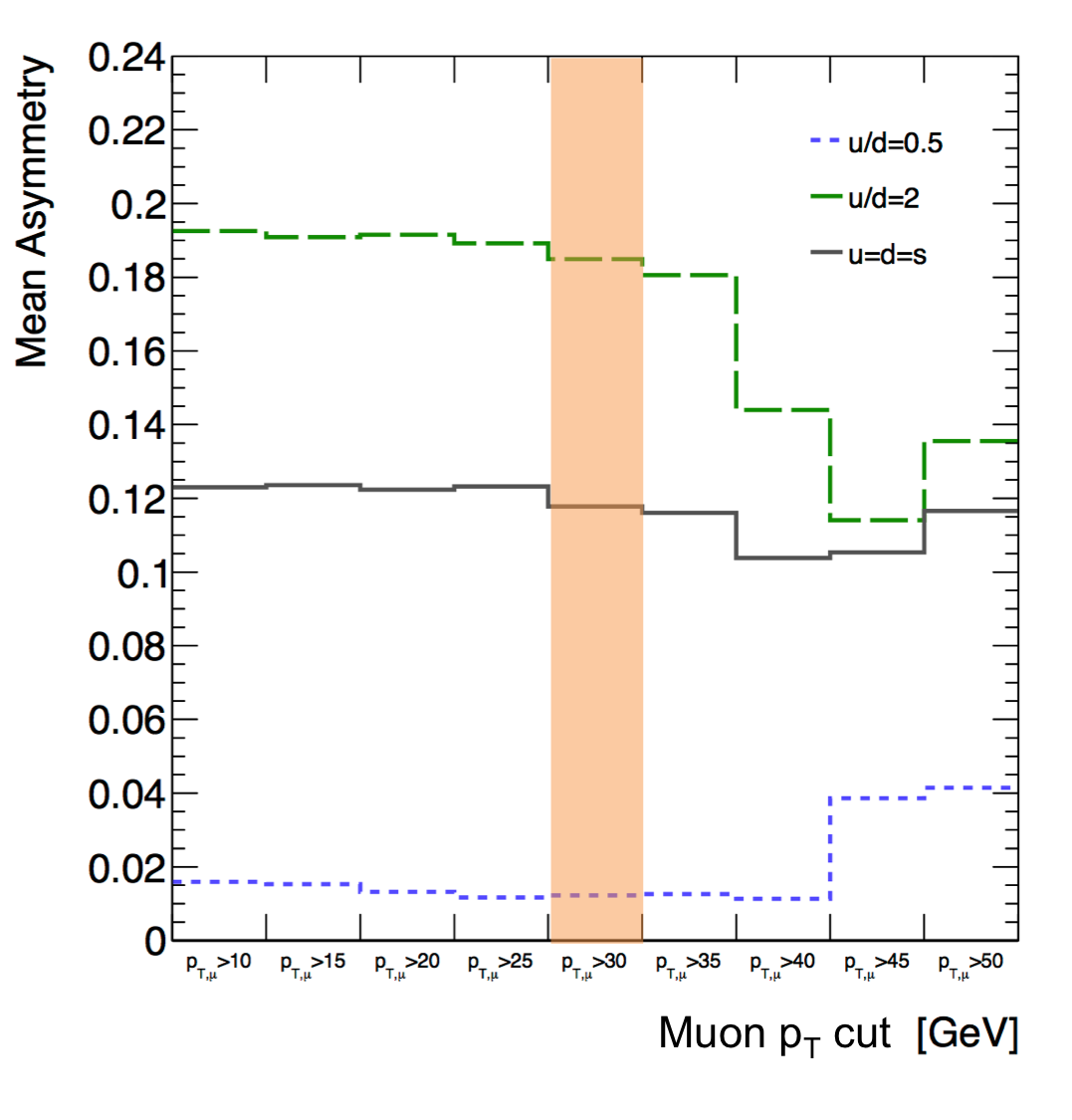}\label{fig:Fig5a}}
\hfill
\subfloat[Missing energy $p_T$ cuts]{\centering \includegraphics[width=.32\linewidth]{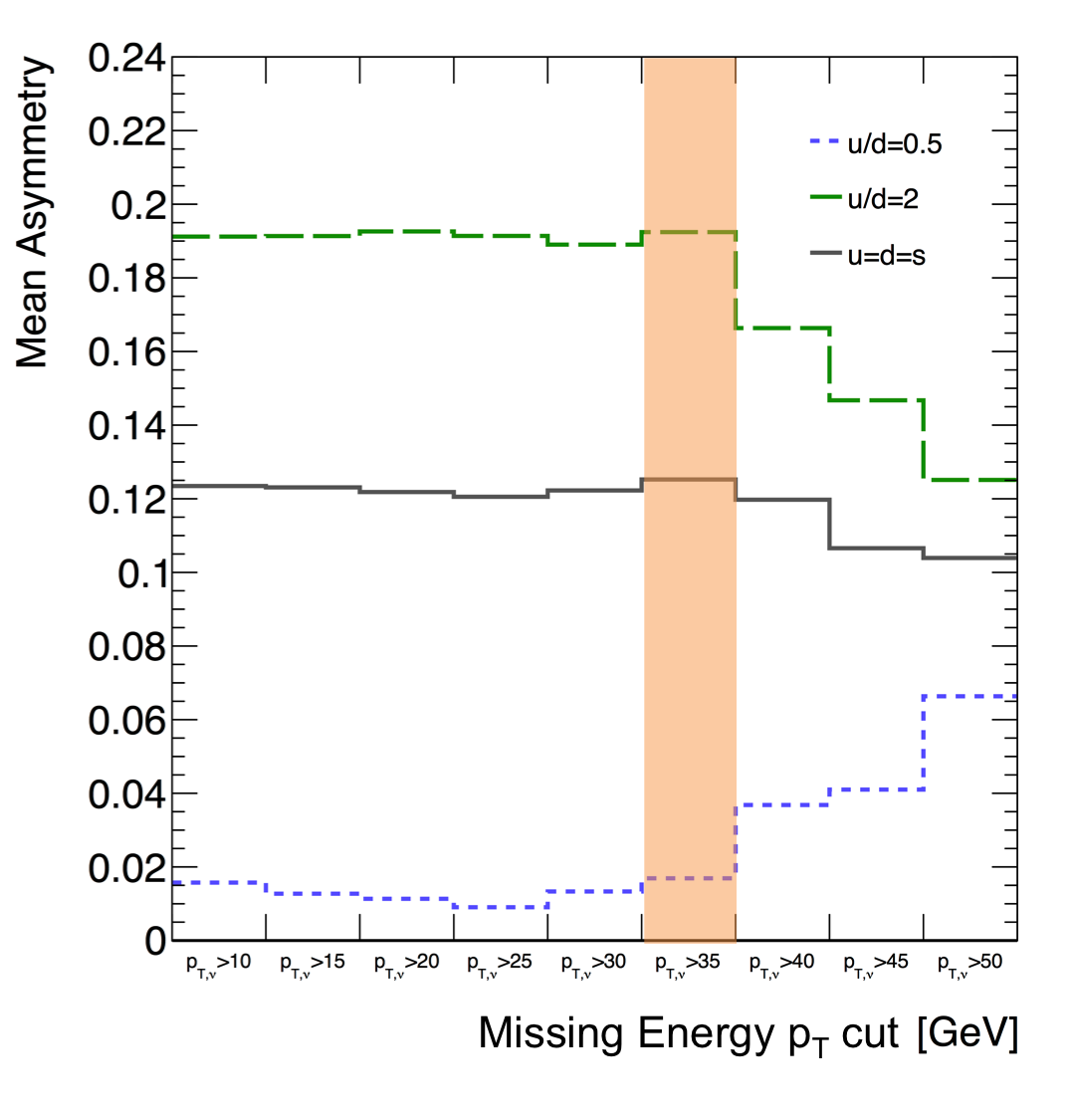}\label{fig:Fig5b}}
\hfill
\subfloat[$m_T(W)$  cuts]{\centering \includegraphics[width=.32\linewidth]{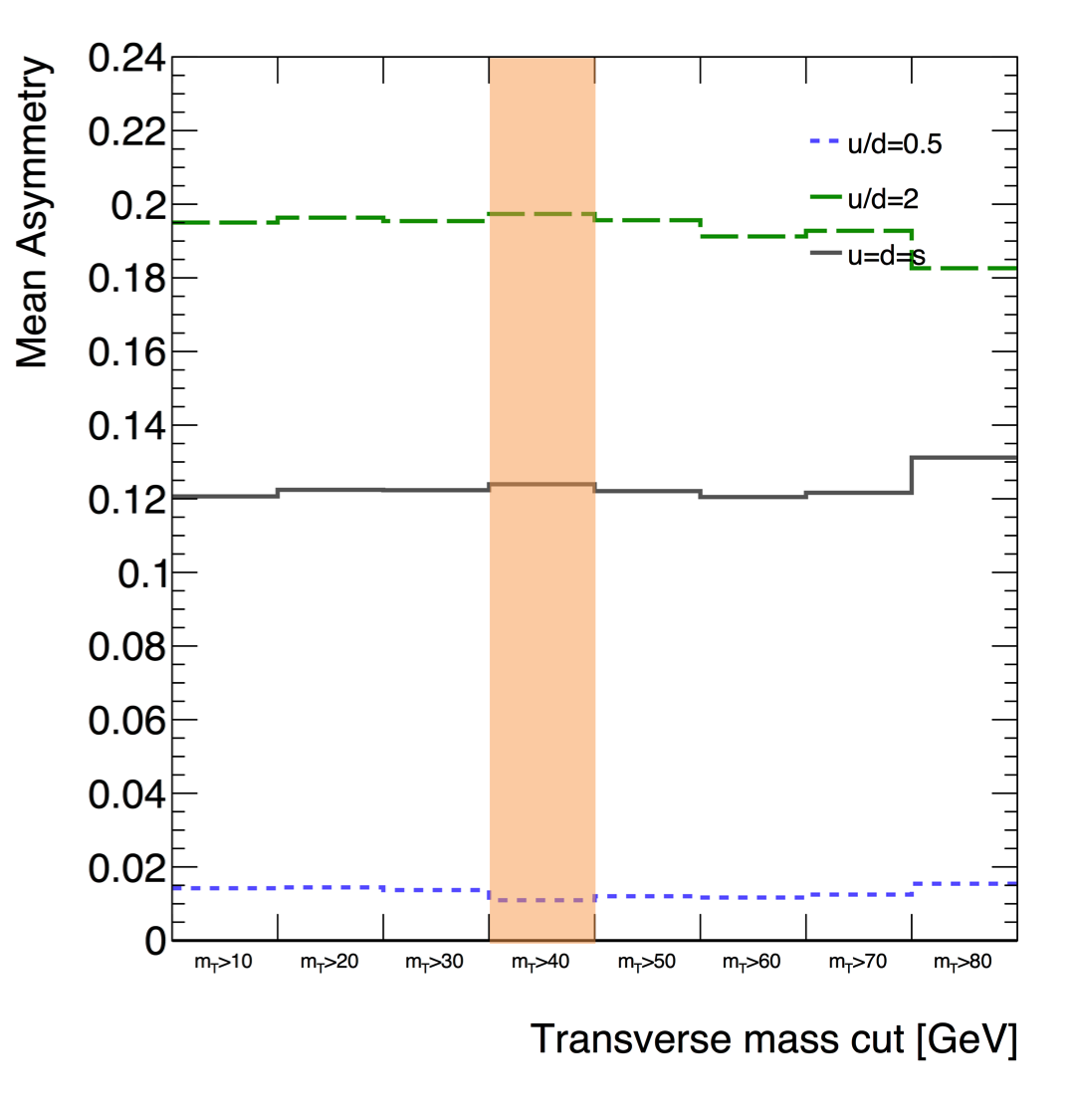}\label{fig:Fig5c}}
\caption{Mean charge asymmetry of single-diffractive W bosons sensitivity to various selection criteria: (a) muon transverse momentum, (b) missing energy associated transverse momentum and (c) transverse mass of the $W$ boson for $\xi<0.12$. The orange shaded bins correspond to the selection applied on the samples. The value of the asymmetry when applying these cuts is compatible with the ones obtained for less stringent requirements.}
\label{fig:Fig5}
\end{figure}

On Fig.\ref{fig:Fig5} we show the sensitivity of the mean asymmetry to various selection cuts on the muon and neutrino $p_T$ and on the $W$ transverse mass.  Above a certain threshold, the mean asymmetry becomes sensitive to the selection criteria applied on the samples. For consistency of the results, we make sure that the mean asymmetry obtained under this requirements lies inside the non-sensitivity domain. The set of selection cuts chosen is represented by shaded orange areas.  We shall notice that the cut on the missing energy associate $p_T$ at 35 GeV is already quite stringent and should not be tightened up.

\section{Conclusions}
\label{sec:sectionV}
The measurement of the single diffractive W boson charge asymmetry using proton tagging at LHC is promising. With enough statistics, it will allow to assess the quarkonic content of the Pomeron and to test the formalism for hadron-hadron scattering. It should be mentioned that the mechanism of diffractive interactions is not yet fully understood. Therefore, measuring a charge asymmetry differing from the expectations would open the door also to other possible scenarios beyond the one covered here.

More than 10 pb$^{-1}$ of luminosity have been predicted to be sufficient  to discriminate between the different scenarios for inclusive measurements. The asymmetry has been proven to be sensitive to the ratio of the diffractive PDFs $u_{\mathbb{P}}$ and $d_{\mathbb{P}}$ for values in the range predicted by HERA. No significant variation was observed for the ratio of $s_{\mathbb{P}}$ and $d_{\mathbb{P}}$. Other assumptions on the DPDFs ratios may be tried that are still compatible with the constraints set by previous HERA measurements. The dependence of $R_{ud}$ to the kinematics can be assessed with 100 pb$^{-1}$ of data. Taking into account the standard selection criteria on the lepton and missing energy products increases the needed luminosity to 300 pb$^{-1}$.

The installation of forward proton detectors located at around 210 m from the ATLAS and CMS interaction points, will provide a larger coverage in $\xi$  than the detectors currently in place, within the region of dominance of the Pomeron amplitude.

Assuming a sufficient luminosity with small pile-up, a measurement of the ratio of Z and W bosons produced in a single-diffractive context, could be performed with the objective of testing the Pomeron content, similarly to what was presented in the analysis featuring dijets and photon+jet~\cite{PhysRevD.88.074029}.

This study is part of the rich diffractive programme at LHC which includes Standard Model tests for the central exclusive production of jets~\cite{ATL-PHYS-PUB-2015-003}, quarkonia~\cite{Altarelli:2014jsa} and vector boson photoproduction~\cite{PhysRevD.90.054003} in addition to the inclusive hard production of jets and photon+jet.  It also extends to more exploratory physics topics, in particular anomalous coupling studies~\cite{Fichet:2014uka, Chapon:2009hh, Kepka:2008yx, PhysRevD.89.114004} probing the existence of extra-dimensions.

\section*{Acknowledgements}

This work is supported in part by the Swiss National Science Foundation Doc.Mobility grant No. P1SKP2\_155196 and by the Polish National Science Centre grant 2012/05/B/ST2\\\noindent/02480.

\newpage
\bibliography{biblio}
\bibliographystyle{JHEP}
\end{document}